\documentclass{article}

\usepackage{amssymb,amsfonts,amsmath}
\usepackage{cite,enumerate,float,indentfirst}
\usepackage{color}

\def\be{\begin{eqnarray}}
\def\ee{\end{eqnarray}}
\def\nn{\nonumber}

\definecolor{red}{rgb}{1,0,0}
\definecolor{orange}{rgb}{1,0.5,0}
\definecolor{violet}{rgb}{0.7,0,1}

\hoffset= - 1.0in         % switch off for draft style
\begin{document}

\title{\vspace{-.1cm}{\Large {\bf q-Painlev\'e equation from Virasoro constraints}\vspace{.2cm}}
\author{
{\bf A.Mironov$^{a,b,c,d}$}\footnote{mironov@lpi.ru; mironov@itep.ru}\ \ and
\ {\bf A.Morozov$^{b,c,d}$}\thanks{morozov@itep.ru}}
\date{ }
}

\maketitle

\vspace{-5.5cm}

\begin{center}
\hfill FIAN/TD-17/17\\
\hfill IITP/TH-13/17\\
\hfill ITEP/TH-22/17
\end{center}

\vspace{3.3cm}

\begin{center}
$^a$ {\small {\it Lebedev Physics Institute, Moscow 119991, Russia}}\\
$^b$ {\small {\it ITEP, Moscow 117218, Russia}}\\
$^c$ {\small {\it Institute for Information Transmission Problems, Moscow 127994, Russia}}\\
$^d$ {\small {\it National Research Nuclear University MEPhI, Moscow 115409, Russia }}
\end{center}

\vspace{.5cm}

\begin{abstract}
The $q$-Painleve equation, satisfied by the Fourier transform
of the $q$-Virasoro conformal blocks at $c=1$,
is interpreted as a reformulation of the string equation and two other
Virasoro constraints in the $5d$ Dotsenko-Fateev matrix model.
\end{abstract}

\bigskip

\bigskip

The rich Seiberg-Witten theory \cite{SWth,SWint} and its quantization \cite{qSW},
involving the Nekrasov functions \cite{Nfunc} and the AGT relations \cite{AGT}
is nowadays understood \cite{AGTmamo,DIM}
as implication of the modern matrix-model theory \cite{UFN3},
applied to the old conformal \cite{confmamo} models,
which originally appeared in the Dotsenko-Fateev (DF) description \cite{DF}
of conformal blocks \cite{BPZ}.
As explained in \cite{MMZ,MMpain}, integrable properties are revealed in the quantities
AGT-related to the Fourier transform of conformal blocks w.r.t. the internal
$\alpha$-parameters, which are also known \cite{GLT} to satisfy the Painlev\'e equations
w.r.t. the cross-ratio of the puncture positions in the conformal block.
In this paper, we claim that these equations are nothing but the ordinary string equations,
provided by the lowest Virasoro constraints in the matrix model,
and this becomes most transparent after the $q$-deformation
AGT-related to the low-energy description of the $5d$ super-Yang-Mills theories.

\bigskip

The DF matrix model at $c=1$ is a matrix model of Penner type \cite{Pen}
with the partition function
\be
Z^{(N)}(\alpha_1,\alpha_2,\alpha_3|z)
= {1\over N!}\int \prod_{i<j}^N (x_i-x_j)^2
\prod_{i=1}^N x_i^{2\alpha_1}(z-x_i)^{2\alpha_2}(1-x_i)^{2\alpha_3}dx_i= \left< \Delta^2(x)\right>
= \nn \\
=\det_{1\leq i,j\leq N}\left< x^{i+j-2}\right>
= \det_{1\leq i,j\leq N} Z^{(1)}\left(\alpha_1+\frac{i+j-2}{2},\alpha_2,\alpha_3\,\Big|\,z\right)
\label{det1}
\ee
The determinant $\det_{1\leq i,j\leq N}\left< x^{i+j-2}\right>$ is equal to $(-1)^{N(N-1)/2}\det_{1\leq i,j\leq N}\left< x^{i-1}(1-x)^{j-1}\right>$, since expanding the latter one into the Newton binomial, one obtains the linear combination of lines in the former determinant. Hence, additionally
\be
Z^{(N)}(\alpha_1,\alpha_2,\alpha_3|z)
= (-1)^{N(N-1)/2}\det_{1\leq i,j\leq N}
Z^{(1)}\left(\alpha_1+\frac{i-1}{2},\alpha_2,\alpha_3+\frac{j-1}{2}\,\Big|\,z\right)
\label{det2}
\ee

\noindent
This partition function is a $\tau$-function of the Toda chain hierarchy
in Miwa representation, with the positions of punctures being the Miwa variables and
the $\alpha$-parameters playing the role of halves of multiplicities \cite{Khar,MMpain}.
In fact, this is still the case for any multiple integral of form (\ref{det1})
with the measure $\prod_{i=1}^Ndx_i\to\prod_{i=1}^Ndf(x_i)$, where $f(x)$
is an arbitrary function.
This function is fixed by an equation additional to the integrable equations,
this is exactly the string equation,
which comes as the first equation(s) of the Virasoro constraints.

Indeed, the partition function also satisfies an infinite set of Virasoro constraints: the Ward identities
associated with the change of integration variables $x_i$ \cite{MMvir}.
It is, however, a non-trivial exercise to express them as operators
acting on the $\alpha$-parameters of $Z^{(N)}$
-- but this can be done and leads to a chain of interesting statements (claims).

\paragraph{\underline{Claim 1}} of the present paper is that for $N=1$
all the Virasoro constraints reduce to just the lowest three,
which act as the multiplicities shifts:
\be
L_{-1}: & \alpha_1 Z^{(1)}\left(\alpha_1-\frac{1}{2},\alpha_2,\alpha_3\right) =
\alpha_2Z^{(1)}\left(\alpha_1,\alpha_2-\frac{1}{2},\alpha_3\right)
+ \alpha_3Z^{(1)}\left(\alpha_1,\alpha_2,\alpha_3-\frac{1}{2}\right)
\nn \\
L_{0}: & \left(\,\alpha_1+\alpha_2+\alpha_3+\underline{\frac{1}{2}}\,\right)
Z^{(1)}\big(\alpha_1,\alpha_2,\alpha_3\big) =
\alpha_2z \cdot Z^{(1)}\left(\alpha_1,\alpha_2-\frac{1}{2},\alpha_3\right)
+ \alpha_3Z^{(1)}\left(\alpha_1,\alpha_2,\alpha_3-\frac{1}{2}\right)
\nn \\
L_{1}: & \left(\alpha_1+\alpha_2+\alpha_3+ { 1} \right)
Z^{(1)}\left(\alpha_1+\frac{1}{2},\alpha_2,\alpha_3\right)
+ (\alpha_2z+\alpha_3) Z^{(1)}\Big(\alpha_1,\alpha_2,\alpha_3\Big) = \nn \\
& = \alpha_2z^2 \cdot Z^{(1)}\left(\alpha_1,\alpha_2-\frac{1}{2},\alpha_3\right)
+ \alpha_3Z^{(1)}\left(\alpha_1,\alpha_2,\alpha_3-\frac{1}{2}\right)
\ee
With the help of the first two the last constraint can be rewritten in a simpler form:
\be
\tilde L_1: \ \ \
\alpha_2(z-1) \cdot \left\{
Z^{(1)}\left(\alpha_1+\frac{1}{2},\alpha_2-\frac{1}{2},\alpha_3\right)
- z\,Z^{(1)}\left(\alpha_1,\alpha_2-\frac{1}{2},\alpha_3\right)
+Z^{(1)}\Big(\alpha_1,\alpha_2,\alpha_3\Big)\right\} = 0
\label{L1hgf}
\ee
Formally, these are examples of Gauss relations \cite{Garel} between hypergeometric functions
with different parameters.

\paragraph{\underline{Claim 2}} is that generalization of
the first two
constraints to $N>1$ (comultiplication) is almost {\it the same}, with just three modifications:

\begin{itemize}

\item $Z^{(1)}$ is substituted by bilinear combination $Z^{(N)}Z^{(N-1)}$,

\item shifts act on the two components in opposite directions, e.g.
$$Z^{(1)}\left(\alpha_1-\frac{1}{2},\alpha_2,\alpha_3\right)
\longrightarrow  Z^{(N)}\left(\alpha_1-\frac{1}{2},\alpha_2,\alpha_3\right)
Z^{(N-1)}\left(\alpha_1+\frac{1}{2},\alpha_2,\alpha_3\right)$$

\item underlined parameter in $L_0$  is slightly changed:
$\frac{1}{2}\longrightarrow N-\frac{1}{2}$.
\end{itemize}

\noindent
These are non-trivially-looking algebraic relations between  hypergeometric functions
at different parameters, which, however, follow from linear Gauss relations.
Derivation of these statements is much simpler if one uses the determinant
formula (\ref{det2}) instead of (\ref{det1}).

\paragraph{\underline{Claim 3}} is that the first two equations involve just
four different functions and are homogeneous, therefore, they imply a
rational relation
\be
w_1 = \frac{\alpha_1}{\alpha_1+\alpha_2+\alpha_3+N-\frac{1}{2}}
\cdot \frac{\alpha_3w_2+\alpha_2 z}{\alpha_3w_2+\alpha_2} \ \ \Longleftrightarrow \ \
w_2 = -\frac{\alpha_2}{\alpha_3}\cdot
\frac{\left(\alpha_1+\alpha_2+\alpha_3+N-\frac{1}{2}\right)w_1-\alpha_1 z}
{\left(\alpha_1+\alpha_2+\alpha_3+N-\frac{1}{2}\right)w_1-\alpha_1}
\label{ratweq}
\ee
between the two ratios:
\be
 w_1 =   \frac{Z^{(N)}\left(\alpha_1,\alpha_2,\alpha_3\right)
 Z^{(N-1)}\left(\alpha_1,\alpha_2,\alpha_3\right)}
 {Z^{(N)}\left(\alpha_1-\frac{1}{2},\alpha_2,\alpha_3\right)
 Z^{(N-1)}\left(\alpha_1+\frac{1}{2},\alpha_2,\alpha_3\right)},
 \ \
 w_2 =  \frac{Z^{(N)}\left(\alpha_1,\alpha_2,\alpha_3-\frac{1}{2}\right)
 Z^{(N-1)}\left(\alpha_1,\alpha_2,\alpha_3+\frac{1}{2}\right)}
 {Z^{(N)}\left(\alpha_1,\alpha_2-\frac{1}{2},\alpha_3\right)
 Z^{(N-1)}\left(\alpha_1,\alpha_2+\frac{1}{2},\alpha_3\right)}
\ee
The third Virasoro constraint then imposes an additional requirement:
\be
\frac{Z^{(1)}\left(\alpha_1+\frac{1}{2},\alpha_2,\alpha_3\right)}
{Z^{(1)}\left(\alpha_1 ,\alpha_2,\alpha_3\right)} +
\frac{Z^{(1)}\left(\alpha_1 ,\alpha_2+\frac{1}{2},\alpha_3\right)}
{Z^{(1)}\left(\alpha_1 ,\alpha_2,\alpha_3\right)} = z
\label{addrec}
\ee

\paragraph{\underline{Claim 4}} is that all this is true for an arbitrary integration contour
in the original matrix integral, in particular, for $Z^{(N)}$ made from the determinants of
\be
Z^{(1)} =  \int_0^z x^{2\alpha_1}(z-x)^{2\alpha_2}(1-x)^{2\alpha_3} dx
+ \mu  \int_1^\infty x^{2\alpha_1}(z-x)^{2\alpha_2}(1-x)^{2\alpha_3} dx
=
\label{2hgf}
\ee
$$
= z^{2\alpha_1+2\alpha_2+1}\int_0^1 x^{2\alpha_1}(1-x)^{2\alpha_2}(1-zx)^{2\alpha_3} dx
+\mu e^{2i\pi( \alpha_2+\alpha_3)}
\int_0^1 x^{-2\alpha_1-2\alpha_2-2\alpha_3-2}(1-x)^{2\alpha_3}(1-zx)^{2\alpha_2} dx
$$
which is a linear combination of two hypergeometric functions with an
arbitrary coefficient $\mu$.
It is inverse Fourier transform in this parameter, which converts $Z^{(N)}$ into the ordinary
4-point conformal blocks.

This is the standard point that the Virasoro constraints as differential
or difference
equations fix solutions only up to a choice of the integration contour \cite{AMM1}.
The space of solutions is then parameterized by the number of independent contours,
sometimes it is just one contour (the Gaussian model),
sometimes there are few (the Dijkgraaf-Vafa solution) \cite{AMM1,Mir,DV}.
For the simplest DF model (\ref{2hgf}),
4-point and $4d$,
there are two independent contours \cite{AGTmamo} and,
hence, the space of solutions is one-dimensional
(since only the relative coefficient between the two contours matters).

\paragraph{\underline{Claim 5}} is that this entire pattern survives a $q$-deformation,
when the matrix model is substituted by the one from \cite{DF5d} with the Jackson integrals
and determinants are made from the combination of $q$-hypergeometric functions:
\be\label{8}
Z_q^{(1)} = z^{2\alpha_1+2\alpha_2+1}\,F_q\big(\alpha_1,\alpha_2,\alpha_3\,\big|\,z\big)
+  \tilde \mu
e^{2\pi i(\alpha_2+\alpha_3)} \,
F_q\big(-\alpha_1-\alpha_2-\alpha_3-1,\alpha_3,\alpha_2\,\big|\,z\big)
\ee
with quantum numbers defined as $[n]=\frac{1-q^n}{1-q}$, Jackson integral as
$\ \int \!f(x)\,d_qx = \sum_{n=0}^\infty\, (1-q)\,q^nf(q^n)\ $ and
\be
F_q(\alpha_1,\alpha_2,\alpha_3|z) =
\int  x^{2\alpha_1}\,(1;x)_{2\alpha_2}\,(1;zx)_{2\alpha_3} \, d_qx =
\ee
{\footnotesize
$$
=
{\frac{\Gamma_q(2\alpha_1+1)\,\Gamma_q(2\alpha_2+1)}{\Gamma_q(2\alpha_1+2\alpha_2+2)}}
\cdot
\sum_{k=0}^\infty \frac{ (-z)^k\,q^{k(k-1)/2}}{[k]! }\,
\frac{\Gamma_q(2\alpha_1+k+1)}{\Gamma_q(2\alpha_1+1)}\,
\frac{\Gamma_q(2\alpha_3+1)}{\Gamma_q(2\alpha_3+1-k)}\,
\frac{\Gamma_q(2\alpha_1+2\alpha_2+2)}{\Gamma_q(2\alpha_1+2\alpha_2+k+2)}=
$$
}
$$
=\mathfrak{B}_q(2\alpha_1+1,2\alpha_2+1)\cdot\!\!\!\!
\phantom{A}_2\phi_1\left(q^{-2\alpha_3},q^{2\alpha_1+1};q^{2\alpha_{1}+2\alpha_2+2}
\,\big|\,q,z\right)
$$
Here $\ (a;q)_p=\prod_{k=0}^{p-1} (1-q^ka)\ $ is the Pochhammer symbol,
$\Gamma_q(x)$ and $\mathfrak{B}_q(x,y)$
are the $q$-Gamma and $q$-Beta-functions respectively, and $\phantom{A}_2\phi_1(a,b;c\,|\,q,z)$
is the Heine basic $q$-hypergeometric function \cite{GasR}.

\noindent
The coefficients in the Virasoro constraints are now quantized and some arguments $z$ are
shifted:

{\footnotesize
\be
L_{-1}: & [2\alpha_1]\, Z_q^{(1)}\left(\alpha_1-\frac{1}{2},\alpha_2,\alpha_3\,\Big|\,z\right)
\ =\ [2\alpha_2]\, Z_q^{(1)}\left(\alpha_1-\frac{1}{2},\alpha_2,\alpha_3\,\Big|\, z\right)
+ [2\alpha_3]\, q^{-2\alpha_2-1}\cdot
Z_q^{(1)}\left(\alpha_1,\alpha_2,\alpha_3-\frac{1}{2}\,\Big|\,qz\right)
\nn \\
L_{0}:\ \ & \!\!\!\!\!\!\! [2\alpha_1+2\alpha_2+2\alpha_3+{1}] \,
Z_q^{(1)}\!\Big(\alpha_1,\alpha_2,\alpha_3\,\Big|\,z\Big) \ = \
[2\alpha_2] \,
q^{2\alpha_1+2\alpha_2+2\alpha_3}
z\, Z_q^{(1)}\!
\left(\alpha_1,\alpha_2-\frac{1}{2},\alpha_3\,\Big|\,\frac{z}{q} \right)
+ [2\alpha_3] \,
Z_q^{(1)}\!\left(\alpha_1,\alpha_2,\alpha_3-\frac{1}{2}\,\Big|\,z\right)
\nn \\
\tilde L_{1}:\ \ &
[2\alpha_2](z-1) \, \left\{ q^{2\alpha_2}\,
Z_q^{(1)}\left(\alpha_1+\frac{1}{2},\alpha_2-\frac{1}{2},\alpha_3\,\Big|\,z\right)
^{\phantom{5}}
- z\, Z_q^{(1)}\left(\alpha_1,\alpha_2-\frac{1}{2},\alpha_3\,\Big|\,z\right)
\ +\ Z_q^{(1)}\Big(\alpha_1,\alpha_2,\alpha_3\,\Big|\,z\Big)\right\} = 0
\nn
\ee
}

\noindent
The new feature of the $N>1$ generalization in this case is that the shifts of $z$
are also made in the opposite directions:
\be
Z^{(1)}\left(\alpha_1-\frac{1}{2},\alpha_2,\alpha_3\,\Big|\,q^s\, z\right)
\longrightarrow  Z^{(N)}\left(\alpha_1-\frac{1}{2},\alpha_2,\alpha_3\,\Big|\,q^s\, z\right)\cdot
Z^{(N-1)}\left(\alpha_1+\frac{1}{2},\alpha_2,\alpha_3\,\Big|\,q^{-s}\,z\right)
\ee
and this makes equation in terms of $w$-variables more complicated.

\paragraph{\underline{Claim 6}} is that
certain manipulations  convert these constraints into
the pair of  $q$-Painleve equations \cite{JS,qP}
\be
{w_1(z)w_1(qz)\over a_3a_4}={(w_2(qz)-b_1z)(w_2(qz)-b_2z)\over (w_2(qz)-b_3)(w_2(qz)-b_4)}\nn\\
{w_2(z)w_2(qz)\over b_3b_4}={(w_1(z)-a_1z)(w_1(z)-a_2z)\over (w_1(z)-a_3)(w_1(z)-a_4)}
\label{PVI}
\ee
where the coefficients $a_p$ and $b_p$ are some powers of $q$ linear in the $\alpha$-parameters,
satisfying
\be
{b_1b_2\over b_3b_4}=q{a_1a_2\over a_3a_4}
\ee
These equations reproduce the rational equations (\ref{ratweq}) in the autonomous $q=1$ limit
and the conventional differential equation Painleve VI in the double scaling
$q\to 1$ limit \cite{JS}.
This is somewhat similar to taking the $q=1$ limit of $f(z)=f(qz)$,
which is a very restrictive $f'(z)=0$ rather than the fully non-constraining $f(z)=f(z)$,
and it is the third Virasoro constraint $L_1$ that plays a role in making the
autonomous limit smooth.

\bigskip

As emphasized in \cite{qP},
%equations (\ref{PVI}) are associated with
the Seiberg-Witten curve for the $SU(2)$ gauge theory
with four matter hypermultiplets
%\be
%{(w_1-a_3)(w_1-a_4)\over w_1}w_2+{b_3b_4(w_1-a_1z)(w_1-a_2z)\over w_1}{1\over w_2}
%-{(b_3+b_4)w_1^2+a_3a_4(b_1+b_2)z\over w_1}= \hbox{const}
%\ee
\be
{(x-a_3)(x-a_4)\over x}\cdot y+{b_3b_4(x-a_1z)(x-a_2z)\over x}\cdot {1\over y}
-{(b_3+b_4)x^2+a_3a_4(b_1+b_2)z\over x}= \hbox{const}
\ee
%which, together with
and the symplectic form $d\log x\wedge d\log y$
are invariant under the transformation
\be
\left(\begin{array}{c} x \\ \\ y \end{array}\right)
= \left(\begin{array}{c} w_1(z) \\ \\ w_2(z) \end{array}\right)
\ \longrightarrow \
\left(\begin{array}{c} X \\ \\ Y \end{array}\right) =
\left(\begin{array}{c} w_1(qz) \\ \\ w_2(qz) \end{array}\right) =
\left(\begin{array}{c} \frac{a_3a_4}{x}\cdot \frac{(Y-zb_1)(Y-zb_2)}{(Y-b_3)(Y-b_4)} \\ \\
\frac{b_3b_4}{y}\cdot\frac{(x-za_1)(x-za_2)}{(x-a_3)(x-a_4)}   \end{array}\right)
\ee
with $a_1a_2b_3b_4=a_3a_4b_1b_2$ inspired by (\ref{PVI}), i.e. in the autonomous limit.
% which is related to (\ref{PVI}) via
%$w_1(z)=x, w_2(z)=y, w_1(qz) = X, w_2(qz)=Y$.
%The same is true for the the symplectic form $d\log x\wedge d\log y$.
This implies the same invariance of the AMM/EO topological recursion \cite{AMM/EO}
and, thus, of the entire non-perturbative DF partition function (conformal blocks).

\bigskip

Thus, we conclude that the $\mu$-deformed partition function of the $5d$ Dotsenko-Fateev
model of \cite{DF5d},
which by the argument of \cite{MMpain} describes the Fourier transform of
the $q$-Virasoro conformal blocks,
satisfies  the $q$-Painleve equations in the form of \cite{JS}, in agreement with \cite{JNS}.
Thus, the $q$-Painleve equation is a direct consequence of the Virasoro constraints
in the matrix model and plays the role of the string equation for the $\tau$-function
in Miwa variables which, in this context, are just the puncture positions
and $\alpha$-parameters are the corresponding multiplicities.
Details of the formulations and evidence in support of them will be provided
in \cite{MMqpainbig}.

\section*{Acknowledgements}

We appreciate discussions with the participants of
the VII Workshop on Geometric Correspondences of Gauge Theories,
especially the explanations and comments by
M.Bershtein, G.Bonelli,  A.Grassi, A.Tanzini, Ya.Yamada and Y.Zenkevich.
This work was performed at the Institute for Information Transmission Problems
with the financial support
of the Russian Science Foundation (Grant No.14-50-00150).

\end{document}